\documentclass{article}
\usepackage{spconf,amsmath,graphicx,hyperref,orcidlink}
\usepackage{amsmath,amssymb,amsfonts}
\usepackage{algorithmic}
\usepackage{graphicx}
\usepackage{textcomp}
\usepackage{xcolor}
\usepackage{url}
\usepackage{comment}
\usepackage{tabularx}
\usepackage{booktabs}
\usepackage{array}
\usepackage{multirow}
\usepackage{tikz}
\usepackage{amssymb}
\usepackage[numbers]{natbib}
\usepackage{setspace}

\title{Brain2Speech-Net: Fast and Intelligible
Brain-to-Speech Synthesis Without Text Decoding}
\name{Shreeram Suresh Chandra\orcidlink{0000-0002-1085-2544}$^{*\dag}$, Zexin Cai\orcidlink{0009-0007-2658-2220}$^*$, Yu Tsao\orcidlink{0000-0001-6956-0418}$^\ddag$, Simon King\orcidlink{0000-0001-6569-0338}$^\S$, Berrak Sisman\orcidlink{0000-0001-8078-3305}$^*$}

\address{$^*$Johns Hopkins University, USA \quad $^\dag$The University of Texas at Dallas, USA \\ $^\ddag$Academia Sinica, Taiwan \quad $^\S$University of Edinburgh, UK}
\begin{document}
%
\maketitle
\begin{abstract}

The loss of speech limits communication for individuals with paralysis. Direct neural-to-speech synthesis is challenging due to the limited availability of neural data for training speech brain–computer interfaces. Most existing systems rely on cascaded neural-to-text-to-speech pipelines, which increase inference latency and propagate errors across stages. We present \textbf{Brain2Speech-Net}, a single-stage neural-to-speech generation framework without intermediate text decoding. We use a differentiable phoneme bottleneck and a deep-HMM alignment mechanism to map long neural recordings into the latent space of a text-to-speech (TTS) model, enabling high-quality speech synthesis. Brain2Speech-Net is the only system in our comparison that produces intelligible speech while generating faster than real time.
\end{abstract}
\vspace{-5mm}
\section{Introduction}
Speech and language are fundamental to human communication \cite{dunbar2009only}.  Decades of neuroscience research have progressively characterized the neural pathways underlying speech perception and production, from classical lesion-based localization \cite{geschwind1970organization} to contemporary dual-stream models of sensorimotor integration. These findings provide the foundation for decoding intended speech directly from brain activity.

Recent breakthroughs on neuron-level speech production understanding, not only provides a deeper understanding of the human brain, but also creates a new form of human communication, one in which we directly communicate our thoughts through neural impulses \cite{moses2021neuroprosthesis}, an idea central to the development of \emph{speech Brain–Computer Interfaces} (speechBCIs). A clinical realization of this vision is the \emph{speech neuroprosthesis} (SN) \cite{jude2025decoding}, designed to restore communication for individuals who have lost the ability to speak due to conditions such as \emph{amyotrophic lateral sclerosis} (ALS). 

Most speechBCIs, including SN, adopt a cascaded paradigm that first decodes text from neural activity and subsequently synthesizes speech using an independent speech generator. While this architecture improves linguistic fluency through strong language model (LM) priors, it introduces important limitations. LMs with strong linguistic priors can improve fluency and grammatical coherence, but may also override neural evidence. Moreover, multi-stage processing pipelines introduces computational latency. Since speechBCIs are intended to function as primary communication devices \cite{brumberg2010brain}, they must operate at human conversational rates without noticeable delays. In practice, such cascaded architectures (1) increase inference latency and (2) propagate compounding errors, limiting responsiveness in real-time communication.

In parallel, speech processing has moved toward unified end-to-end architectures in tasks such as automatic speech recognition (ASR) \cite{graves2012sequence} and speech-to-speech translation (SST) \cite{jia2022translatotron}, reducing latency and mitigating compounding errors. These developments motivate us to explore whether a structured single-stage neural-to-speech model can balance intelligibility and latency in speechBCIs. However, unlike ASR and SST, speechBCIs are an emerging field with very limited data to train large-vocabulary models, making fully end-to-end training particularly challenging.

\begin{figure*}[t]
    \centering
    \scalebox{0.35}
    {\includegraphics{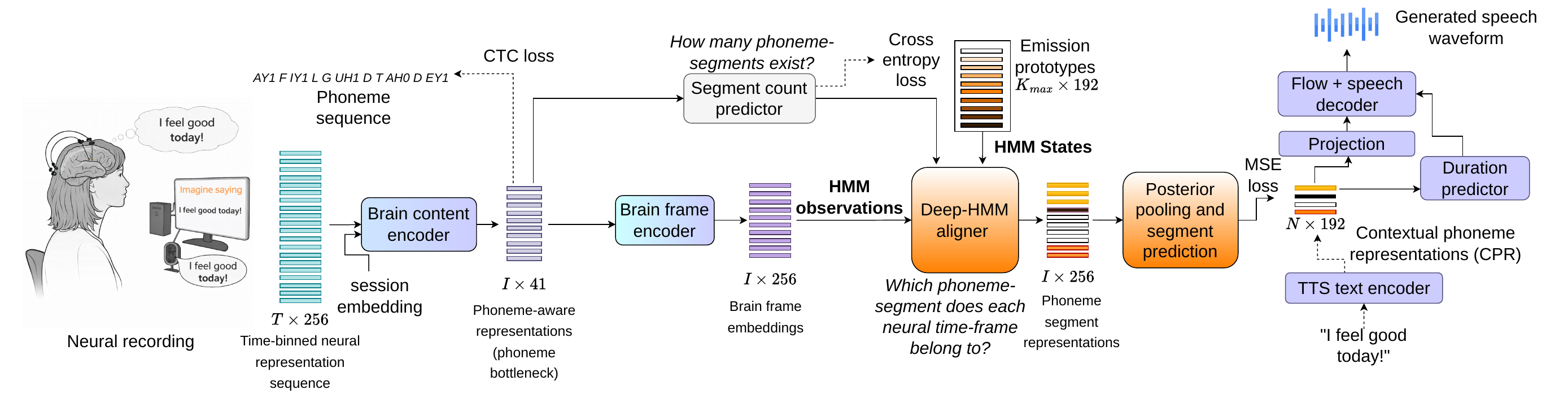}}
    \vspace{-6mm}
    \caption{Brain2Speech-Net architecture for single-stage neural-to-speech synthesis using a phoneme-informed bottleneck and a deep-HMM alignment to VITS encoder representations. The solid lines represent the data flow path during both train and inference, the dotted lines are data paths used only during training
}
    \label{framework}
\end{figure*}

In this paper, we introduce a phoneme-informed bottleneck that bridges a brain content encoder and a speech decoder, enabling single-stage inference for neural-to-speech synthesis. The bottleneck leverages frame-level phoneme logits as an intermediate linguistic representation, providing linguistic guidance while avoiding explicit text decoding. To support data-efficient alignment between neural activity and \emph{contextual phoneme representations} (CPRs) from the TTS text encoder, we incorporate a lightweight and differentiable hidden Markov model alignment module, with transitions and emissions parameterized by neural networks. 
The main contributions of this work are as follows: 
\textbf{(1)} We introduce \textbf{Brain2Speech-Net}, a linguistically-informed neural-to-speech architecture that enables single-stage speech synthesis from intracortical brain signals, without intermediate text decoding. \textbf{(2)} We incorporate phoneme-aware representations as a differentiable linguistic bottleneck, preserving phonemic structure while avoiding explicit text generation. \textbf{(3)} We develop a lightweight differentiable deep-HMM that learns monotonic alignment between long neural recordings and contextual phoneme representations in a pre-trained TTS latent space, leveraging decoder acoustic priors for intelligible speech generation. \textbf{(4)} Brain2Speech-Net enables intelligible real-time speech, whereas cascaded models introduce latency and direct-speech-unit models remain largely unintelligible in limited-data settings. To support reproducibility, we release code and pretrained models of Brain2Speech-Net\footnote{\scriptsize{Brain2Speech-Net is available at \url{https://b2s-lang.github.io/}}}.

\vspace{-3mm}

\section{Related work}

\textbf{Brain-to-text decoding.} Advances in neural recording technologies have enabled decoding of speech and language representations directly from brain activity \cite{wang2024decoding}. Research in neurolinguistics has identified neural correlates of speech generation across motor intentions ranging from imagined to overt speech \cite{tang2023semantic,tang2024imagined}. Prior studies have explored multiple recording modalities, including electrocorticography (ECoG) \cite{angrick2019speech}, fMRI \cite{almufareh2025inner}, MEG \cite{kwon2024direct}, and fNIRS \cite{cooney2021bimodal}, demonstrating that neural signals encode phonetic, lexical, and articulatory information. More recently, single-neuron recordings using micro-electrode arrays (MEA) in motor cortex have revealed fine-grained phonetic representations \cite{willett2023high}. Building on these findings, the Brain-to-Text benchmark \cite{willett2024brain} gave rise to many competitive spoken content decoding methods. Notably, much of the performance improvement in these models has been shown to emerge from language modeling rather than improvements in neural signal extraction \cite{li2025brain}. \\
\textbf{Brain-to-speech decoding.} Closely related to text decoding is the direct synthesis of speech from neural signals. Most speech neuroprostheses first decode text and then render it with a text-to-speech system, though recent work has explored bypassing text entirely \cite{littlejohn2025streaming}. Metzger et al.~\cite{metzger2023high} map ECoG features to frame-level discrete speech units with a bidirectional RNN and resynthesize them with a neural vocoder, reaching roughly $25\%$ WER on a 1024-word vocabulary. BrainTalker \citep{kim2023braintalker} addresses the low-resource nature of brain-to-speech synthesis via transfer learning over self-supervised speech representations, while Wairagkar et al.~\cite{wairagkar2025instantaneous} show that intracortical activity carries prosodic information and argue for preserving paralinguistic cues in the generated speech. We build on this direction, emphasizing direct synthesis as a path to low-latency neural communication.


\vspace{-3mm}
\section{Intracortical Data}

We use the intracortical speech neuroprosthesis dataset of Willett et al.~\cite{willett2023high}, recorded from a participant with ALS during prompted attempted speech. Neural activity from four microelectrode arrays is binned into threshold crossings and spike band power across 512 channels. Following prior work, we use the 256 motor-cortex channels. The participant cannot produce intelligible speech, so no aligned acoustic recordings exist. We therefore synthesize pseudo-speech targets from the prompted text with a pre-trained VITS model~\cite{kim2021conditional}, giving 8,800 training and 880 test utterances (5.71 hours).


\vspace{-2mm}
\section{Brain2Speech-Net}
\begingroup
\setlength{\abovedisplayskip}{2pt}
\setlength{\belowdisplayskip}{2pt}
\setlength{\abovedisplayshortskip}{1pt}
\setlength{\belowdisplayshortskip}{1pt}
\setlength{\jot}{1pt}
\vspace{-3pt}

Brain2Speech-Net is a single-stage neural-to-speech model that uses phoneme-aware representations as a differentiable linguistic bottleneck, as shown in Figure \ref{framework}. Intracortical recordings contain phonetic information but are limited in scale, while pre-trained TTS models provide strong acoustic priors from large speech corpora. By predicting phoneme-aware representations from neural signals, the model preserves neural evidence while aligning efficiently to \emph{contextual phoneme representations} (CPRs) in the TTS text-encoder latent space. Brain2Speech-Net is trained in two phases. In Phase 1, the brain content encoder is pre-trained to produce phoneme-aware representations from neural activity. In Phase 2, the full model, including the brain content encoder, is trained jointly to synthesize speech. 
At inference, speech is generated in one pass: neural signals $\rightarrow$ CPRs $\rightarrow$ waveform - without explicit text decoding.


\textbf{Phase 1: Pre-training the brain content encoder.} The brain content encoder takes time-binned neural features ($\mathbf{x}_{0:T-1}$) and the session-specific embedding ($\mathbf{e}_{\mathrm{session}}$) as input and produces frame-level representations $\mathbf{n}_i \in \mathbb{R}^{d_{bc}}$. 
The encoder is trained using the connectionist temporal classification (CTC) loss \citep{graves2006connectionist}, with the ground-truth phoneme sequence ($\mathbf{y}$) as supervision. 
\[
\mathcal{L}_{\mathrm{CTC}}
=
-\log p\!\left(\mathbf{y} \mid \mathbf{x}_{0:T-1}, \mathbf{e}_{\mathrm{session}}\right)
\]


The CTC objective enables learning frame-level phoneme-aware representations without requiring explicit alignment between neural frames and phoneme targets. To account for distribution shifts across recording sessions, we incorporate a learned session embedding that conditions the encoder on the recording session \citep{feng2024towards}. After Phase~1, the encoder produces phoneme-aware representations that serve as input to the alignment module in Phase~2.

\textbf{Phase 2: Deep-HMM alignment.} 
Next, we align the long neural representation sequence $\mathbf{x}_{0:T-1}$ with the much shorter sequence of contextual phoneme representations (CPRs) $\mathbf{z}_{0:N-1}$ from the VITS text encoder. Following neural-network parameterized HMMs for data-efficient monotonic alignment \citep{ mehta2022neural}, we adopt a left-to-right, no-skip \emph{hidden Markov model} (HMM) over neural frames and phoneme segments. Each state indexes a phoneme segment position in the target CPR sequence rather than a phoneme class, giving segment-level alignment under limited data. We take CPR targets from the VITS text encoder because it parametrizes a conditional prior over latent speech representations during TTS training, providing a probabilistically well-defined and acoustically robust target space. The brain content encoder first maps $\mathbf{x}_{0:T-1}$ to phoneme-aware representations $\mathbf{n}_{0:I-1}$, which the brain frame encoder then maps to the frame embedding space $\mathbf{e}_i$. The brain content encoder is fine-tuned throughout this phase.

\[
\begin{aligned}
\mathbf{n}_{0:I-1} &= \mathrm{BrainContentEnc}_{\psi}(\mathbf{x}_{0:T-1}, \mathbf{e}_{\mathrm{session}}), \\
\mathbf{e}_i &= \mathrm{BrainFrameEnc}_{\theta}(\mathbf{n}_i).
\end{aligned}
\]

The sequence $\mathbf{e}_{0:I-1}$ forms the observation sequence for the deep-HMM aligner, which assigns each neural time frame to a corresponding phoneme segment in the CPR sequence. \\
\textbf{States and emissions.} Let $k \in \{0, \ldots, N-1, N_{\mathrm{EOS}}\}$ index phoneme segments, where $N_{\mathrm{EOS}}$ is a self-absorbing end state. Each state has an embedding $\mathbf{v}_k$ drawn from a learned lookup table. For each segment $k$, we define a reference vector $\mathbf{c}_k \in \mathbb{R}^{d_{state}}$. During training, $\mathbf{c}_k$ is obtained by projecting the ground-truth CPR latent 
$\mathbf{z}_k \in \mathbb{R}^{d_{\mathrm{cpr}}}$:
\[
\mathbf{c}_k = \mathbf{W}_{\mathrm{tgt}} \mathbf{z}_k,
\qquad
\log E_i(k) = - \left\lVert \mathbf{A}\mathbf{e}_i - \mathbf{c}_k \right\rVert^2 .
\]
where $\mathbf{W}_{\mathrm{tgt}} \in \mathbb{R}^{d_{state} \times d_{\mathrm{cpr}}}$ is learned, $\mathbf{A} \in \mathbb{R}^{d_{state} \times d_f}$ projects frame embeddings, and $E_i(k)$ is the emission score for state $k$ at index $i$. During inference, ground-truth CPRs are unavailable. Instead, $\mathbf{c}_k$ is obtained from a learned emission prototype table $\mathbf{P} \in \mathbb{R}^{K_{\max} \times d_{state}}$ as $\mathbf{c}_k = \mathbf{P}[k]$, where $K_{\max}$ is the maximum alignment states per utterance.

\noindent\textbf{Transitions.} At index $i$, state $k$ may remain in $k$, advance to $k{+}1$, or transition to $\mathrm{EOS}$. Transition probabilities are predicted by a neural network $h_\theta$:
\[
\tau^{\mathrm{adv}}_{i,k}, \; \tau^{\mathrm{stop}}_{i,k}
= h_\theta(\mathbf{e}_i, \mathbf{v}_k),
\qquad
\tau^{\mathrm{stay}}_{i,k}
= 1 - \tau^{\mathrm{adv}}_{i,k} - \tau^{\mathrm{stop}}_{i,k}.
\]
The model is constrained to enter $\mathrm{EOS}$ only from the final segment $N{-}1$.

\noindent\textbf{Forward--backward inference.} Forward–backward inference computes posterior state occupancies $\gamma_{i,k}$ using forward variables $\alpha_i(k)$, which accumulate evidence from past observations up to index $i$, and backward variables $\beta_i(k)$, which accumulate evidence from future observations until termination at $\mathsf{EOS}$~\citep{Rabiner89-ATO}. These posteriors define a soft alignment that is used to pool frame embeddings into phoneme segment representations.
The posterior of segment $k$ at index $i$ is
\[
\gamma_{i,k}
=
\frac{\alpha_i(k)\beta_i(k)}
{\sum_j \alpha_i(j)\beta_i(j)}.
\]

\noindent\textbf{Posterior pooling and segment prediction.}  The frame embeddings are passed to the \emph{Posterior-Pooled Segment Predictor} (PPSP), which first performs $\gamma_{i,k}$-weighted pooling from the deep-HMM to produce a sequence aligned in length with the VITS encoder targets and then predicts CPRs $\hat{\mathbf{z}}$. Training minimizes a \emph{mean squared error} (MSE) loss between $\hat{\mathbf{z}}$ and the ground-truth latents $\mathbf{z}$ over valid segments, with gradients propagated through the soft alignments. \\
\[
\hat{\mathbf{z}}_{\mathrm{k}}
=
\mathrm{PPSP}_{\phi}\ \left(
\sum_{i=0}^{I-1} \gamma_{i,k}\,\mathbf{e}_i
\right),
\qquad k = 0,\ldots,N-1
\]

\[
\mathcal{L}_{\mathrm{MSE}}
=
\frac{1}{N} \sum_{k=0}^{N-1}
\left\lVert \hat{\mathbf{z}}_{k} - \mathbf{z}_{k} \right\rVert^2,
\qquad
\mathcal{L}_{\mathrm{len}}
=
- \log p(N \mid \mathbf{n}_{0:I-1}).
\]
where $N$ is the number of valid phoneme segments and $k$ are the indices of the phoneme segments. 


Since $N$ is unknown at inference, an auxiliary segment count predictor is trained with cross-entropy loss ($\mathcal{L}_{len}$). A CTC loss on a linear head over the brain content encoder features preserves the linguistic bottleneck. Gradients from both losses are detached from the alignment objective and optimized independently. At inference, the predicted $N$ defines the HMM state sequence; emissions use learned prototypes, and predicted CPRs drive the VITS decoder for single-stage neural-to-speech synthesis.

\endgroup
\vspace{-3mm}
\section{Experiments}

\textbf{Model and training details.} The brain content encoder is a 5-layer GRU over windowed neural features (kernel $32$, stride $2$), yielding phoneme-aware representations $\mathbf{n}_i \in \mathbb{R}^{41}$, and is fine-tuned jointly with the Phase~2 alignment components at a lower learning rate ($10^{-5}$). A linear layer followed by three residual MLP blocks (dropout $0.1$) maps $\mathbf{n}_i$ to frame embeddings $\mathbf{e}_i \in \mathbb{R}^{256}$. The segment count predictor is a three-layer MLP ($256 \rightarrow 128 \rightarrow K_{\max}$, ReLU, dropout $0.1$) over an attention-pooled summary of $\{\mathbf{n}_i\}$; transitions use a two-hidden-layer MLP ($512 \rightarrow 256$). The PPSP applies posterior-weighted pooling, a two-layer segment head, a four-layer Transformer encoder ($d{=}192$, 4 heads, FFN $768$), and a residual refinement MLP. Emission prototypes $\mathbf{P} \in \mathbb{R}^{K_{\max} \times 192}$ are initialized by $k$-means ($k{=}K_{\max}$, 10 restarts) on projected training CPR latents $\{\mathbf{W}_{\mathrm{tgt}}\mathbf{z}_k\}$ and re-clustered every 5 epochs. We set $\lambda_{\mathrm{CTC}}{=}0.1$ and optimize with Adam (batch size $32$, peak learning rate $10^{-3}$, 500-step warm-up).

\noindent\textbf{Baselines.} We compare against cascaded brain-to-text-to-speech (NTS) pipelines and a direct neural-to-speech baseline. We use the same brain content encoder across the board. \emph{NTS-Cascade (5-gram)} \citep{willett2024brain} decodes the phoneme-aware representations from the brain content encoder to text using a 5-gram language model (125k-word vocabulary trained on Switchboard text) with beam width 18, followed by VITS speech synthesis. \emph{NTS-Cascade (5-gram + LLM re-rank)} extends this pipeline by re-ranking a 100-best list from the 5-gram decoder using a large language model (OPT-8B) before speech synthesis. \emph{Direct-DSU synthesis} \citep{metzger2023high} predicts HuBERT discrete speech units \citep{hsu2021hubert} directly from neural features using a bidirectional RNN and generates waveforms with a neural vocoder \citep{van2022comparison}. \\
\noindent\textbf{Evaluation metrics.}We report phoneme error rate (PER) from Wav2Vec~2.0-Large-LV60 \citep{baevski2020wav2vec} fine-tuned for phoneme recognition, word error rate (WER) from Whisper-medium \citep{radford2023robust}, mel-cepstral distortion (MCD) \citep{kubichek1993mel} after DTW alignment, and real-time factor (RTF). Cascaded baselines additionally report WER of the decoded text against the ground-truth transcript (\textit{Text} WER). We conduct subjective MOS listening tests \citep{wells2024experimental}.
\vspace{-4mm}
\section{Results}
\vspace{-7mm}
\begin{table}[h]
\centering
\setlength{\tabcolsep}{3pt}
\renewcommand{\arraystretch}{1.0}
\caption{Unseen-utterance (open-set) evaluation comparing cascaded and single-stage speech BCI models. WER and PER are reported as fractions and MCD is reported in dB.}
\resizebox{\columnwidth}{!}{%
\begin{tabular}{lccccc}
\toprule
\textbf{Model}
& \textit{Text} WER$\downarrow$ & PER$\downarrow$ & WER$\downarrow$ & MCD$\downarrow$ & RTF$\downarrow$ \\
\midrule
NTS-Cascade (5-gram) \citep{willett2024brain}
& 0.197 & 0.209 & 0.229 & 4.967 & 1.846 \\
NTS-Cascade (5-gram + LLM) \citep{willett2024brain}
& 0.172 & 0.188 & 0.206 & 4.894 & 1.962 \\
\midrule
DSU baseline \citep{metzger2023high}
& \textemdash & 0.785 & 1.054 & 6.181 & 0.060 \\
\textbf{Brain2Speech-Net}
& \textemdash & 0.373 & 0.472 & 5.522 & 0.532 \\
\bottomrule
\end{tabular}%
} \\
\vspace{3mm}
  \textit{Ground-truth speech reference:} WER$_\text{held out}$ = 0.025.
\label{tab:results_open}
\end{table}

\begin{figure}[t]
\centering
\scalebox{0.8}{
\includegraphics[width=\linewidth]{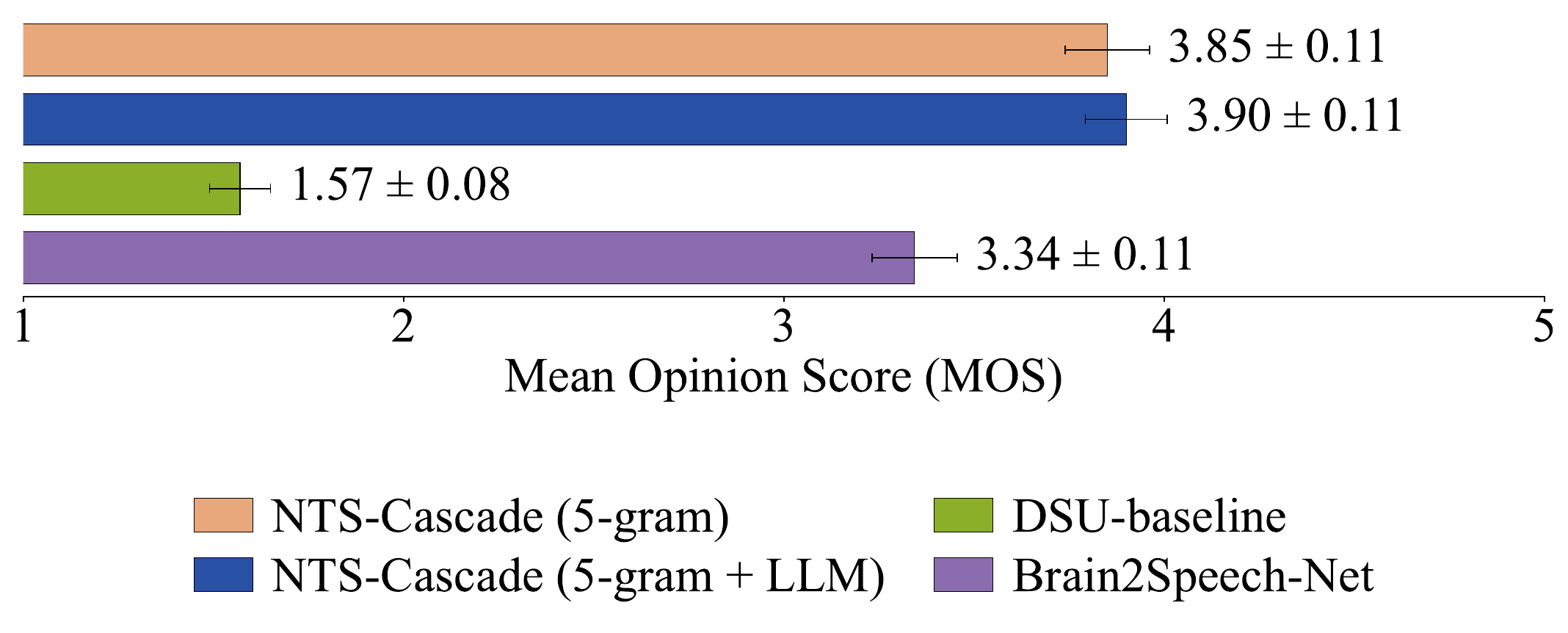}
}
\vspace{-5mm}
\caption{Subjective speech intelligibility measured via mean opinion score (MOS) by human listeners for the evaluated models; error bars show 95\% confidence intervals.}

\label{fig:mos}
\vspace{-6mm}
\end{figure}
\vspace{-3mm}
\textbf{Objective Evaluation.} Table~\ref{tab:results_open} reports results on the official test split. The
cascaded pipelines attain the lowest error rates: NTS-Cascade (5-gram) reaches a
WER of 0.229, and LLM re-ranking lowers it to 0.206. Both decode through an
explicit 125k-word language model. Brain2Speech-Net operates without such a
lexical constraint and reaches a PER of 0.373 and a WER of 0.472, improving over
the DSU baseline by 0.412 PER and 0.582 WER under the same brain content encoder.
The differentiable phoneme interface therefore retains phonemic structure that
acoustic-only decoding does not recover. Brain2Speech-Net generates at an RTF of
0.53, against above 1.8 for the cascades; DSU is faster ($\sim$0.06) but less
intelligible. Brain2Speech-Net thus gives the best accuracy--speed trade-off
among the compared systems. Ground-truth speech reaches a WER of 0.025 on these
utterances, and the residual gap reflects the difficulty of open-vocabulary
neural-to-speech mapping from 5.2 hours of single-participant intracortical data.\\
\textbf{Listening Experiments.} We conducted an MOS intelligibility study with 20 English-speaking listeners on unseen utterances. Each system contributed 20 samples (80 total), presented in randomized order; listeners rated intelligibility on a 1--5 scale (1 = poor, 5 = excellent). As shown in Figure. \ref{fig:mos}, Brain2Speech-Net achieves an MOS of 3.34, indicating that listeners found the speech intelligible. \\
\textbf{Compounding Errors.} Both cascades lose accuracy at the TTS stage: text WER rises from 0.197 to 0.229 for NTS-Cascade (5-gram) and from 0.172 to 0.206 with LLM.\\
\noindent\textbf{Training fit as a diagnostic.}
To probe the future direction of this work, we ask whether the model struggles with (a) learning the neural-to-speech mapping or (b) generalizing to unseen utterances. These failures are confounded in the unseen test set. Hence, we repeat the evaluation on 880 training utterances sampled uniformly across recording sessions, whose neural trials and sentence prompts are both seen during optimization (Table~\ref{tab:results_closed}; arrows denote held-out $\rightarrow$ seen). Brain2Speech-Net moves 0.472 $\rightarrow$ 0.068 WER and 0.373 $\rightarrow$ 0.109 PER, against 0.229 $\rightarrow$ 0.059 WER for NTS-Cascade (5-gram) and 0.025 $\rightarrow$ 0.021 for the ground-truth speech reference. On seen data the single-stage model is within 0.009 WER of the cascade, so the phoneme bottleneck and the alignment to contextual phoneme representations are sufficient to express the mapping. The held-out gap is a failure of generalization, not of capacity. The DSU baseline moves 1.054 $\rightarrow$ 1.011 WER and 0.785 $\rightarrow$ 0.758 PER, so direct acoustic decoding does not fit the training data at this scale.
\begin{table}[h]
\centering
\setlength{\tabcolsep}{3pt}
\renewcommand{\arraystretch}{1.0}
\vspace{-2mm}
\caption{Training-fit evaluation on 880 utterances drawn from the training set.
WER and PER are fractions, MCD is in dB}
\resizebox{\columnwidth}{!}{%
\begin{tabular}{lccccc}
\toprule
\textbf{Model}
& \textit{Text} WER$\downarrow$ & PER$\downarrow$ & WER$\downarrow$ & MCD$\downarrow$ & RTF$\downarrow$ \\
\midrule
NTS-Cascade (5-gram) \citep{willett2024brain}
& 0.037 & 0.078 & 0.059 & 4.415 & 1.664 \\
NTS-Cascade (5-gram + LLM) \citep{willett2024brain}
& 0.021 & 0.069 & 0.042 & 4.363 & 1.714 \\
\midrule
DSU baseline \citep{metzger2023high}
& \textemdash & 0.758 & 1.011 & 6.004 & 0.056 \\
\textbf{Brain2Speech-Net}
& \textemdash & 0.109 & 0.068 & 4.632 & 0.504 \\
\bottomrule
\end{tabular}%
} \\
\vspace{3mm}
  \textit{Ground-truth speech reference:} WER$_\text{seen}$ = 0.021.
\label{tab:results_closed}
\end{table}

\vspace{-5mm}
\section{Conclusion}
We presented Brain2Speech-Net, a single-stage neural-to-speech framework that removes intermediate text decoding. A differentiable phoneme bottleneck and a deep-HMM alignment module map long intracortical recordings into the latent space of a pre-trained TTS model, enabling data-efficient training under limited neural data. Brain2Speech-Net is the only system in our comparison that produces intelligible speech while generating faster than real time; extending this to unseen utterances remains the open problem.

{\footnotesize
    \setstretch{0.85}
\bibliographystyle{IEEEbib}
\bibliography{strings,refs}}

\end{document}